\documentclass{article}
\usepackage{spconf,amsmath,graphicx,hyperref}

\usepackage{tabularx}
\usepackage{booktabs}
\usepackage{amssymb}

\usepackage{pifont}
\usepackage{enumitem}
\usepackage{fontawesome5}

\usepackage{xcolor}
\usepackage{threeparttable}
\usepackage{siunitx} 

\usepackage{multirow}

\definecolor{mygreen}{RGB}{0, 186, 19}
\definecolor{myred}{RGB}{214, 0, 11}

\title{Decoding the Ear: A Framework for Objectifying Expressiveness from Human Preference Through Efficient Alignment}

\name{
    Zhiyu Lin$^{1, 2}$\thanks{Email: 224040288@link.cuhk.edu.cn},
    Jingwen Yang$^{1, 2}$,
    Jiale Zhao$^{2}$,
    Meng Liu$^{2}$,
    Sunzhu Li$^{2}$,
    Benyou Wang$^{1\ast}$\thanks{$^{\ast}$Email: wangbenyou@cuhk.edu.cn}
}
\address{
    $^{1}$The Chinese University of Hong Kong, Shenzhen, China \\
    $^{2}$Li Auto Inc., China
}

\begin{document}
\ninept
\maketitle
\begin{abstract}

Recent speech-to-speech (S2S) models generate intelligible speech but still lack natural expressiveness, largely due to the absence of a reliable evaluation metric. Existing approaches, such as subjective MOS ratings, low-level acoustic features, and emotion recognition are costly, limited, or incomplete. To address this, we present \textbf{DeEAR} (\textbf{De}coding the \textbf{E}xpressive Preference of e\textbf{AR}), a framework that converts human preference for speech expressiveness into an objective score. Grounded in phonetics and psychology, DeEAR evaluates speech across three dimensions: Emotion, Prosody, and Spontaneity, achieving strong alignment with human perception (Spearman's Rank Correlation Coefficient, SRCC = 0.86) using fewer than 500 annotated samples. Beyond reliable scoring, DeEAR enables fair benchmarking and targeted data curation. It not only distinguishes expressiveness gaps across S2S models but also selects 14K expressive utterances to form \textit{\textbf{ExpressiveSpeech}}, which improves the expressive score (from 2.0 to 23.4 on a 100-point scale) of S2S models. Demos and codes are available at \url{https://github.com/FreedomIntelligence/ExpressiveSpeech}
\end{abstract}


\section{Introduction}
\label{sec:intro}

Recent end-to-end speech models can generate clear speech in Text-to-Speech (TTS) tasks. Yet in conversational settings, their output \textbf{often sounds robotic} and lacks the expressiveness vital for applications such as voice assistants, role-playing, and AI companions. The core of this problem is the absence of a reliable evaluation metric.

While fields like speech recognition and audio enhancement benefit from WER~\cite{ali2018wer,morris2004wer} and DNSMOS~\cite{reddy2021dnsmos}, expressiveness still relies on subjective MOS~\cite{le2024limits}, which is costly and unscalable. Existing alternatives are limited: low-level acoustic features~\cite{sisman2020overview} (e.g., pitch, energy) miss perceptual subtleties, and emotion recognition~\cite{straulino2023missing} captures only one facet of expressiveness. In short, \textbf{a comprehensive, human-aligned metric is urgently needed.}

To address this gap, we introduce \textbf{DeEAR}, a novel framework that transforms human preference for speech expressiveness into a reliable, objective score. Building on established theories in phonetics (e.g., intonational phonology~\cite{ladd2008intonational}) and psychology (e.g., the circumplex model of affect~\cite{russell1980circumplex}), we define expressiveness along three core dimensions: \textbf{Emotion}, \textbf{Prosody}, and \textbf{Spontaneity}. We then train a \textbf{unified model} to capture these dimensions and align them with human preference, producing a single expressiveness score. Notably, our method \textbf{achieves a Spearman correlation of 0.86 with human perception using fewer than 500 annotated samples}, making it both data-efficient and practically scalable.

To demonstrate its utility, we apply DeEAR in two key tasks. First, DeEAR provides \textbf{a reliable and convenient framework} for quantifying speech expressiveness through objective scores. It demonstrates strong consistency, achieving a high correlation with human rankings of systems (SRCC = 0.96), and exhibits strong discriminative power. For example, when comparing state-of-the-art dialogue systems, the gap between the highest-scoring (DouBao) and lowest-scoring (Qwen2.5-Omni) models reaches 60.1 points.

Second, DeEAR can also be \textbf{used to curate data}, selecting highly expressive speech to support the training of more expressive TTS or S2S models. In practice, we applied DeEAR to several open-source datasets with potential expressiveness (e.g., Expresso~\cite{nguyen2023expresso}, NCSSD~\cite{liu2024generative}), using a threshold of 63.5 to extract approximately 14K utterances, named \textbf{\emph{ExpressiveSpeech}}. We then \textbf{fine-tuned an S2S model} with this curated dataset, which led to a substantial improvement in expressiveness: the overall expressiveness score rose from 2.0 to 23.4. All three sub-dimensions improved, with particularly notable gains in emotion (from 5.7 to 15.9) and spontaneity (from 33.7 to 62.0).

\section{DeEAR}
\label{sec:method}

This section introduces the methodology of \textbf{DeEAR}. Scoring an abstract concept like \textit{expressiveness} with a single model is unreliable due to limited training data. To address this, we follow four principles: (1) decompose the expressiveness into concrete, solvable tasks; (2) design specialized models for each task to ensure accuracy; (3) align outputs with human preference using limited but interpretable data; and (4) enhance efficiency and scalability. These principles are ultimately instantiated in a four-stage pipeline (Figure~\ref{fig:pipeline} (A)).

\begin{figure}[h]
  \centering
  \includegraphics[width=\columnwidth]{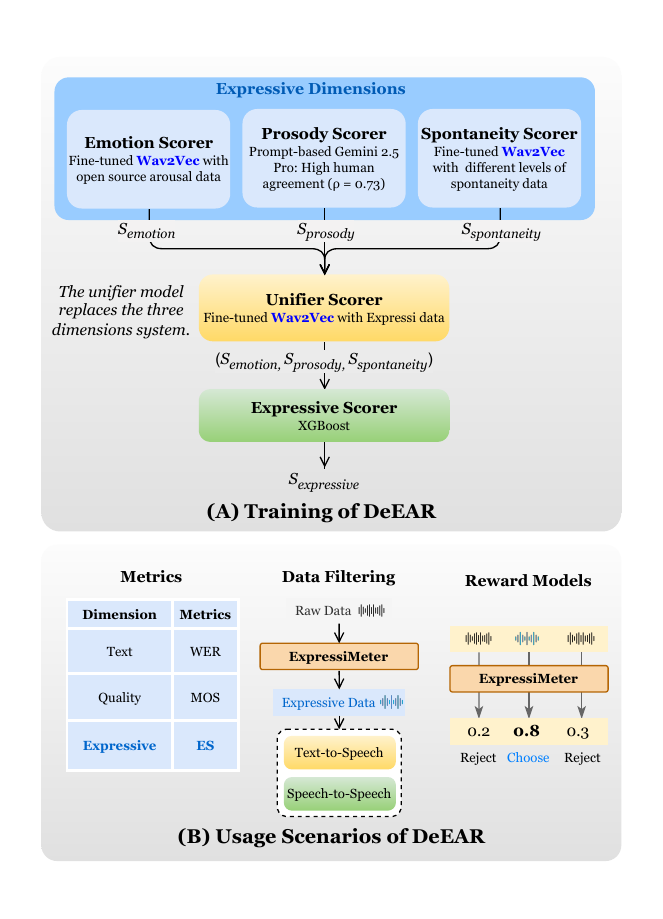} 
\label{fig:pipeline}
\vspace{-10pt}

\caption{The DeEAR Framework: 
  (A) The training follows a four-stage pipeline: Stage\,1 decomposes expressiveness into \textit{Emotion}, \textit{Prosody}, and \textit{Spontaneity}; Stage\,2 trains a scorer for each dimension; Stage\,3 learns a \textit{Unifier Scorer} from these dimension scores; and Stage\,4 trains XGBoost to produce the final expressiveness score $S_{\text{expressive}}$. 
  (B) Applications include filtering audio to build high-quality datasets and serving as a reward model to select outputs from generative models. Here, ES denotes the \textit{Expressiveness Score}.}

  \label{fig:pipeline}
\end{figure}

\subsection{Decomposing Expressiveness for Alignment}
\label{ssec:decomposition}

Drawing from linguistics, psychology, and computational paralinguistics~\cite{schuller2013computational,scherer2003vocal,russell1980circumplex}, we decompose \textit{expressiveness} into three complementary dimensions. (i) Emotion intensity, a central element of expressiveness~\cite{scherer2003vocal}, is measured by arousal~\cite{russell1980circumplex} and correlates with acoustic cues such as pitch range and intensity~\cite{banse1996acoustic}. (ii) Prosody, the melody and rhythm of speech, is fundamental to expressiveness as it conveys the speaker's attitudes and intentions beyond the literal words~\cite{ladd2008intonational,gussenhoven2004phonology,prieto2015intonational}. (iii) Spontaneity reflects perceived authenticity, which listeners infer from acoustic cues such as disfluencies and variable prosody~\cite{shriberg2005spontaneous,de2015information}. While theoretically distinct, the three components interact in human perception, leading us to design a modular system with a final fusion layer to combine them.

\subsection{Proxy Modeling for Sub-Dimensions}
\label{ssec:proxy}

We adopted a task-specific approach to measure each dimension, training specialized models where applicable. Specifically, we used supervised learning for the data-rich Emotion Intensity; leveraged a large language model to generate labels for the subjective and data-scarce Prosodic Richness; and applied a hybrid rule-based heuristic for Spontaneity.

\subsubsection{Emotion Intensity Scoring}

Emotion intensity, defined here as arousal, is a well-established paralinguistic construct. The availability of large labeled datasets makes it well-suited to supervised learning. We therefore fine-tuned the state-of-the-art \texttt{wav2vec2-large-\allowbreak robust-\allowbreak 12-\allowbreak ft-\allowbreak emotion-\allowbreak msp-\allowbreak dim} model, which was already pre-trained for emotion recognition on English speech data. To improve bilingual performance, we further trained it on 12,000 Chinese samples from CNSCED~\cite{wu2025cnsced} and 2,000 English samples from IEMOCAP~\cite{busso2008iemocap}.

\subsubsection{Prosodic Richness Scoring}

Evaluating prosodic richness is challenging because it is subjective and lacks a large-scale labeled dataset. Traditional acoustic features often fail to distinguish engaging from unpleasant melodies.

To overcome these issues, we used LMMs as proxies for human perception, drawing on their ability to judge expressive qualities directly from audio. Using carefully engineered prompts, Gemini 2.5 Pro served as an automated annotator for prosodic quality. Its scores achieved a strong Spearman’s rank correlation (SRCC=0.73) with human ratings, validating the approach and enabling scalable generation of consistent prosodic richness scores ($\boldsymbol{S_{\text{pros}}}$).

\subsubsection{Spontaneity Scoring}

Our scoring of spontaneity is based on the premise that perceived spontaneity (sounding unscripted) requires perceived naturalness (sounding human). This is similar to the speech uncanny valley effect~\cite{Lameris2023Prosody}, where technically perfect voices sound unnatural because they lack human-like imperfections~\cite{matsunaga2022improving}. Recent studies confirm that this loss of naturalness also reduces perceived spontaneity~\cite{Li2023Towards}. We call the cause of this problem perceptual incongruence: a mismatch between high acoustic quality and a non-human speech style.

We employ a two-stage, knowledge-guided supervised strategy.

\textbf{Stage 1: Heuristic-Based Pseudo-Label Generation.} We designed a heuristic function that combines a categorical base level of spontaneity, $L_{\text{base}} \in \{1, 3, 5, 7, 9\}$, with an acoustic quality metric, $M_{\text{avg}}$. The base level is manually assigned at the dataset level. This metric is the mean of four DNSMOS outputs (OVRL, SIG, BAK, P.808 MOS)~\cite{reddy2021dnsmos}. The score is calculated conditionally:
\begin{equation}
S_{\text{spon}} = 
\begin{cases} 
  \text{map}_{\text{penalty}}(M_{\text{avg}}) & \text{if \textit{hyper-clean} and } L_{\text{base}} < L_{\text{max}} \\
  \text{map}_{\text{normal}}(M_{\text{avg}}) & \text{otherwise}
\end{cases}
\label{eq:spontaneity}
\end{equation}
A sample is considered \textit{hyper-clean} when all four underlying DNSMOS metrics exceed a threshold $T_q=3.5$. The $\text{map}_{\text{normal}}(\cdot)$ function linearly scales $M_{\text{avg}}$ to a target range (e.g., $[L_{\text{base}}-1, L_{\text{base}}+1]$), rewarding quality for congruent cases. In contrast, the $\text{map}_{\text{penalty}}(\cdot)$ function performs a reverse linear scaling to a much narrower, predefined punitive range (e.g., $[0.0, 0.5]$ for $L_{\text{base}}=1$). This aggressively penalizes perceptually incongruent samples that are too clean for their category.

\textbf{Stage 2: Supervised Model Fine-tuning.} We then used these pseudo-labels to fine-tune the same \texttt{wav2vec2-large-robust} model backbone used for emotion scoring. This process distills our explicit, knowledge-based heuristic into a robust deep learning model, creating the final spontaneity scorer.

\subsection{Learning the Human Preference Fusion Function}
\label{ssec:fusion}
The core of our alignment strategy lies in an explicit fusion function, engineered to model the complex, non-linear mapping from our sub-dimension scores to a holistic human judgment. This function is designed as a separate, lightweight module to ensure interpretability and fidelity to human preference data.

To model this, we collected a small dataset of 480 audio clips, for which three human annotators provided a single, overall expressiveness score. Using the three proxy scores ($\boldsymbol{S_{\text{emo}}}$, $\boldsymbol{S_{\text{pros}}}$, $\boldsymbol{S_{\text{spon}}}$) as input features, we trained a XGBoost model~\cite{chen2016xgboost} to predict the human-annotated overall score. The resulting model serves as our preference fusion function, capable of predicting a holistic expressiveness score by learning the complex interplay and non-linear trade-offs between the sub-dimensions directly from human preference data.

\subsection{Distillation and Decoupling for a Modular System}
\label{ssec:distillation}

To convert the powerful but demanding \textit{teacher} system for deployment, we employ a twofold strategy: knowledge distillation for efficiency and architecture decoupling for interpretability.

In the distillation step, the capabilities of the three proxy models are compressed into a single student model, \textbf{DeEAR-Base}. The teacher system is applied to 20{,}000 unlabeled utterances to produce pseudo-labels for $\boldsymbol{S_{\text{emo}}}$, $\boldsymbol{S_{\text{pros}}}$, and $\boldsymbol{S_{\text{spon}}}$. DeEAR-Base adopts a \texttt{wav2vec2-large-xlsr-53}~\cite{conneau2020unsupervised} backbone with three regression heads, jointly trained in a multi-task setup to predict the sub-dimensions, thus inheriting nuanced perceptual capabilities in a significantly more efficient form.

In the decoupling step, the final overall score $\boldsymbol{S_{\text{expr}}}$ is not generated directly by DeEAR-Base; instead, its sub-scores are passed to an independently trained XGBoost fusion layer (Section~\ref{ssec:fusion}). This modular design makes the preference logic explicit and detachable, allowing future updates without expensive retraining of the backbone. The combination of DeEAR-Base and the fusion layer constitutes the final \textbf{DeEAR}, which not only yields an objective expressiveness score but also supports practical uses such as filtering training data and guiding generative models (Figure~\ref{fig:pipeline} (B)). For clarity, all scores from DeEAR—overall expressiveness ($\boldsymbol{S_{\text{expr}}}$) and the sub-dimensions of Emotion ($\boldsymbol{S_{\text{emo}}}$), Prosody ($\boldsymbol{S_{\text{pros}}}$), and Spontaneity ($\boldsymbol{S_{\text{spon}}}$)—are presented on a 0-100 scale, where higher is better.

\section{High-Expressive Bilingual Dataset}
\label{sec:dataset_construction}

Existing dialogue datasets often lack consistent vocal expressiveness. To address this gap, we developed \textbf{\emph{ExpressiveSpeech}}, a real world dataset built specifically for high-quality, expressive speech.

The dataset contains approximately 14,000 utterances, totaling 51 hours, with a Chinese-English language ratio close to 1:1. It is composed of curated samples from five open-source emotional dialogue datasets: Expresso~\cite{nguyen2023expresso}, NCSSD~\cite{liu2024generative}, M$^3$ED~\cite{zhao2022m3ed}, MultiDialog~\cite{park2024let}, and IEMOCAP~\cite{busso2008iemocap}. Our pipeline ensures that all selected data meets high standards for both acoustic quality and expressiveness. As shown in Table~\ref{tab:comparasion}, our dataset achieves a significantly higher average expressiveness score of \textbf{80.2} compared to its sources.

\begin{table}[h]
  \centering
  \setlength{\tabcolsep}{4pt}
  \caption{Comparison of ExpressiveSpeech with its source datasets. $\boldsymbol{L_\text{expr}}$ marks datasets with explicit expressiveness labels. $\boldsymbol{S_{\text{expr}}}$ is the average expressiveness scored from our DeEAR, with the highest score highlighted in bold.}
  \label{tab:comparasion}
  \begin{tabular}{lccccc}
    \toprule
    \textbf{Dataset} & \textbf{Language} & \textbf{Duration(h)}  & $\boldsymbol{L_\text{expr}}$ & $\boldsymbol{S_{\text{expr}}}$ \\
    \midrule
    Multidialog~\cite{park2024let} & EN & 340 & \ding{55} & 39.4 \\
    M$^3$ED~\cite{zhao2022m3ed} & ZH & 14 &   \ding{55} & 49.9 \\
    NCSSD~\cite{liu2024generative} & EN, ZH & 236 &  \ding{55} & 50.1 \\
    IEMOCAP~\cite{busso2008iemocap} & EN & 12  &  \ding{55} & 50.9 \\
    Expresso~\cite{nguyen2023expresso} & EN & 46 &  \ding{55} & 62.9 \\
    \hline
    ExpressiveSpeech & EN, ZH & 51 & \checkmark & \textbf{80.2} \\
    \bottomrule
  \end{tabular}
\end{table}

\subsection{Data Curation Pipeline}
Our curation pipeline consists of four main stages to ensure quality.

\textbf{Standardization and Enhancement:} We first standardized all audio to 16kHz mono and segmented multi-turn dialogues into single-speaker utterances. We used ClearerVoice~\cite{zhao2025clearervoice} to remove background noise and separate overlapping speakers. This process significantly improved audio clarity.

\textbf{Quality and Expressiveness Scoring:} We evaluated overall speech quality using DNSMOS P.835 OVRL score, achieving an average of \textbf{3.17}. For expressiveness, we used DeEAR to assign scores to each utterance based on its Emotion, Prosody, and Spontaneity.

\textbf{High-Expressiveness Subset Selection:} We set an expressiveness score threshold of \textbf{63.5} to select the final dataset. This value was determined empirically to align with human perception of high expressiveness. The threshold effectively selects samples that humans perceive as highly expressive and filters out utterances with low or unclear expressiveness.

\textbf{Metadata Organization:} Finally, we generated text transcriptions for 
audio samples using Automatic Speech Recognition (ASR).

\subsection{Ethical Considerations and Licensing}
The construction of ExpressiveSpeech adhered to strict ethical guidelines. It is derived from public, anonymized academic datasets containing no personally identifiable information (PII), and we followed all original data protocols. In line with the non-commercial restrictions of its sources, the dataset is released under the CC BY-NC-SA 4.0 license.

% V2-0910
\section{Experiments}
\label{sec:exp}

\subsection{Validity: Alignment with Human Perception}
\label{ssec:validation}

\textit{DeEAR demonstrates a strong alignment with human perception of expressiveness.}
To validate this, we created four test sets, each containing 100 utterances. 
These sets were composed of diverse audio, including real-world conversations, professional recordings, and TTS-generated speech.

We then asked three graduate students in speech processing to independently rate each utterance on a 1-to-5 scale. 
The ratings followed a standardized protocol with clear definitions and anchor examples. 
The human judgments showed strong reliability, achieving a Krippendorff's alpha of $\boldsymbol{\alpha = 0.72}$. 
We averaged these ratings to create the final ground-truth score for our evaluation.

As shown in Table~\ref{tab:correlation}, DeEAR's scores strongly correlate with the human ratings. 
For the overall expressiveness score ($\boldsymbol{S_{\text{expr}}}$), our metric achieved a Pearson Correlation Coefficient (PCC) of \textbf{0.91} and a Spearman's Rank Correlation Coefficient (SRCC) of \textbf{0.86}. 
These high correlations provide compelling evidence that DeEAR accurately quantifies speech expressiveness.

\begin{table}[h]
  \centering

  \caption{Correlation between DeEAR scores and human ratings. Pearson (PCC) and Spearman (SRCC) coefficients are reported for three dimensions (Emotion, Prosody, Spontaneity) and the overall expressiveness score.}
  \label{tab:correlation}
  \begin{tabular}{lcc}
    \toprule
    \textbf{Dimension} & \textbf{PCC} & \textbf{SRCC} \\
    \midrule
    Emotion ($\boldsymbol{S_{\text{emo}}}$) & 0.72 & 0.65 \\
    Prosody ($\boldsymbol{S_{\text{pros}}}$) & 0.70 & 0.68 \\
    Spontaneity ($\boldsymbol{S_{\text{spon}}}$) & 0.84 & 0.84 \\
    \hline
    Expressiveness ($\boldsymbol{S_{\text{expr}}}$) & \textbf{0.91} & \textbf{0.86} \\
    \bottomrule
  \end{tabular}
\end{table}

\subsection{Application 1: Automated Benchmarking of SOTA Models}
\label{ssec:benchmark}

\textit{DeEAR enables reliable automated model benchmarking, achieving a near-perfect rank correlation (SRCC) of \textbf{0.96} with human evaluations.}
This capability addresses a critical need in the field, as benchmarking state-of-the-art (SOTA) models is vital for progress but is often limited by slow, expensive, and subjective listening tests.

To demonstrate this utility, we used DeEAR to rank seven leading S2S models, including both open- and closed-source systems. For a fair comparison, each model generated a response for the same 20 audio prompts, which covered a range of conversational emotions. We then compared the automated ranking with that from human listeners. This human ranking was created by four native speakers who rated each model's output on a 3-point MOS scale.

The results in Table~\ref{tab:ranking} quantitatively substantiate our claim. 
Beyond the near-perfect rank correlation, the metric also demonstrates strong discriminative power, creating a wide overall score gap of nearly 60 points between the top and bottom-performing systems. 
This confirms that DeEAR can reliably replace manual evaluations for system-level model comparison, providing a scalable and objective solution to a key challenge in speech synthesis research.

\begin{table}[ht]
  \caption{Automated benchmarking of SOTA models using DeEAR versus human evaluation. The rankings demonstrate a near-perfect align (SRCC = 0.96). The table presents scores for overall expressiveness ($\boldsymbol{S_{\text{expr}}}$) and its sub-dimensions, with final ranks in parentheses. \textcolor{mygreen}{Green} and \textcolor{myred}{Red} in the ranks indicate that the DeEAR rank is better or worse than the human rank, respectively.}
  \label{tab:ranking}
  \setlength{\tabcolsep}{4pt}
  \begin{tabularx}{\columnwidth}{
      X
      S[table-format=2.1]
      S[table-format=2.1]
      S[table-format=2.1]
      S[table-format=2.1, table-space-text-post={(9)}]
      S[table-format=2.1, table-space-text-post={(9)}]
    }
      \toprule
      \multirow{2}{*}{\textbf{Model}} & \multicolumn{4}{c}{\textbf{DeEAR Scores}} & {\multirow{2}{*}{\textbf{Human}}} \\
      \cmidrule(lr){2-5}
      & $\boldsymbol{S_{\text{emo}}}$ & $\boldsymbol{S_{\text{pros}}}$ & $\boldsymbol{S_{\text{spon}}}$ & $\boldsymbol{S_{\text{expr}}}$ & \\
      \midrule
      {Doubao}        & \textbf{67.7} & \textbf{58.6} & \textbf{92.5} & \textbf{65.4} {$(1)$} & \textbf{84.2} {$(1)$} \\
      Grok-4 Voice    & 64.8 & 51.7 & 76.8 & 45.2 {$(2)$} & 80.8 {$(2)$} \\
      GPT-4o Audio    & 56.2 & 39.4 & 67.4 & 31.1 {$(\textcolor{myred}{4})$} & 66.3 {$(3)$} \\
      Sesame          & 40.9 & 33.2 & 88.4 & 44.9 {$(\textcolor{mygreen}{3})$} & 56.1 {$(4)$} \\
      Step Audio 2   & 44.2 & 34.3 & 69.4 & 29.3 {$(5)$} & 42.9 {$(5)$} \\
      Qwen2.5-Omni    & 44.4 & 37.6 & 31.9 &  5.3 {$(\textcolor{myred}{7})$} & 41.2 {$(6)$} \\ 
      Gemini-2.5 Pro  & 39.5 & 30.3 & 40.1 &  7.0 {$(\textcolor{mygreen}{6})$} & 34.7 {$(7)$} \\
      \bottomrule
    \end{tabularx}
\end{table}

\subsection{Application 2: Evaluation-driven Data Curation}
\label{ssec:application}

Having established DeEAR as a valid metric and benchmark, we demonstrate its utility in an evaluation-driven paradigm. We aim to prove that a reliable metric can guide data curation to systematically improve a model's expressive capability.

\subsubsection{Experimental Design}
\label{sssec:app_setup}

To quantify the contribution of our method to high-quality data curation, we performed SFT on our S2S model \textbf{Expressive-FT (Ours)} with \textbf{ExpressiveSpeech} mentioned in Section \ref{sec:dataset_construction}. This model, analogous to architectures like MinMo~\cite{chen2025minmo} and Qwen2.5-Omni~\cite{xu2025qwen2}, integrates a 7B LLM with a 1.5B (audio language model) ALM and employs the S3tokenizer. The ALM underwent 230,000 hours of pre-training followed by 4,000 hours of post-training. The complete model was then fine-tuned on the 51-hour \textbf{ExpressiveSpeech} dataset for a single epoch at a learning rate of 1e-5. Both models were assessed on a 100-utterance test set, partitioned into \textbf{in-domain} (held-out from source corpora) and \textbf{out-of-domain} (from unseen sources like Emilia) data to test generalization. The evaluation involved objective scoring with DeEAR and a subjective A/B preference test with 10 native speakers, who chose the more expressive output or declared a tie.

\subsubsection{Results and Analysis}
\label{sssec:app_results}

\textit{DeEAR successfully guides data curation, yielding a model of superior expressiveness.}

\textbf{Objective Results}:
As shown in Table~\ref{tab:objective_results}, our model significantly outperforms the baseline across all dimensions. The model's strong generalization, evidenced by the minimal performance drop on out-of-domain data, stems from its gains being concentrated on highly transferable emotion and spontaneity cues. Our curation process prioritized these dimensions as they were the most significant deficiencies, leading to less focus on the comparatively higher-scoring baseline for prosody. T-tests confirmed that all reported gains are statistically significant ($p < 0.001$), underscoring the efficacy of our targeted data curation for both familiar and unseen data distributions.

\textbf{Subjective Results}: Human evaluations corroborated these findings. In A/B preference tests, listeners favored our Expressive-FT model in \textbf{78.5\%} of cases, versus just \textbf{10\%} for the baseline, with \textbf{11.5\%} rated as ties. This strong preference is statistically significant ($p < 0.001$), providing ground-truth validation of our model's superior expressiveness.

The strong agreement between DeEAR's objective scores and human preference provides conclusive evidence for our central thesis: a powerful, human-aligned metric is the key to systematically and effectively developing more expressive conversational AI.

\begin{table}[h]
  \centering
  \setlength{\tabcolsep}{3.5pt} 

  \caption{Objective results for in-domain, out-of-domain, and overall test sets. The proposed Expressive-FT model consistently outperforms the baseline across expressiveness ($\boldsymbol{S_{\text{expr}}}$), emotion ($\boldsymbol{S_{\text{emo}}}$), prosody ($\boldsymbol{S_{\text{pros}}}$), and spontaneity ($\boldsymbol{S_{\text{spon}}}$), with all gains statistically significant ($p < 0.001$).}
  \label{tab:objective_results}
  \begin{tabular}{llllll}
    \toprule
    \textbf{Set} & \textbf{Model}  & $\boldsymbol{S_{\text{emo}}}$ & $\boldsymbol{S_{\text{pros}}}$ & $\boldsymbol{S_{\text{spon}}}$ & $\boldsymbol{S_{\text{expr}}}$ \\
    \midrule
    \multirow{2}{*}{In-domain} & Baseline  & 5.9 & 35.6 & 34.1 & 2.3\\
    & \textbf{Ours}  & \textbf{15.8} & \textbf{35.8} & \textbf{62.9} & \textbf{24.0}\\
    \midrule
    \multirow{2}{*}{Out-of-domain} & Baseline  & 5.4 & 36.3 & 33.2 & 1.8\\
    & \textbf{Ours} & \textbf{15.9} & \textbf{37.6} & \textbf{61.1} & \textbf{23.0} \\
    \midrule
    \multirow{2}{*}{\textbf{Overall}} & Baseline  & 5.7 & 35.7 & 33.7 & 2.0\\
    & \textbf{Ours} & \textbf{15.9} & \textbf{36.7} & \textbf{62.0} & \textbf{23.4} \\
    \bottomrule
  \end{tabular}
\end{table}

\section{Conclusion}

In this paper, we introduced DeEAR, a human-aligned and data-efficient metric for multi-dimensional speech expressiveness. By capturing Emotion, Prosody, and Spontaneity, DeEAR achieves strong correlation with human perception and scales beyond costly subjective evaluation. Leveraging this metric, we curated ExpressiveSpeech, a large-scale bilingual dataset of highly expressive speech, and fine-tuned a baseline S2S model to achieve substantial improvements in expressiveness. Our findings establish a paradigm of evaluation-driven data curation, underscoring that reliable metrics are crucial for advancing expressive speech synthesis. Future directions include extending DeEAR to reinforcement learning for end-to-end expressiveness optimization.

\bibliographystyle{IEEEbib}
\bibliography{refs}

\begin{thebibliography}{10}

\bibitem{ali2018wer}
Ahmed Ali and Steve Renals,
\newblock ``Word error rate estimation for speech recognition: e-wer,''
\newblock in {\em Proceedings of the 56th Annual Meeting of the Association for Computational Linguistics (Volume 2: Short Papers)}. Association for Computational Linguistics (ACL), 2018, pp. 20--24.

\bibitem{morris2004wer}
Andrew~Cameron Morris, Viktoria Maier, and Phil~D Green,
\newblock ``From wer and ril to mer and wil: improved evaluation measures for connected speech recognition.,''
\newblock in {\em Interspeech}, 2004, pp. 2765--2768.

\bibitem{reddy2021dnsmos}
Chandan~KA Reddy, Vishak Gopal, and Ross Cutler,
\newblock ``Dnsmos: A non-intrusive perceptual objective speech quality metric to evaluate noise suppressors,''
\newblock in {\em ICASSP 2021-2021 IEEE International Conference on Acoustics, Speech and Signal Processing (ICASSP)}. IEEE, 2021, pp. 6493--6497.

\bibitem{le2024limits}
S{\'e}bastien Le~Maguer, Simon King, and Naomi Harte,
\newblock ``The limits of the mean opinion score for speech synthesis evaluation,''
\newblock {\em Computer Speech \& Language}, vol. 84, pp. 101577, 2024.

\bibitem{sisman2020overview}
Berrak Sisman, Junichi Yamagishi, Simon King, and Haizhou Li,
\newblock ``An overview of voice conversion and its challenges: From statistical modeling to deep learning,''
\newblock {\em IEEE/ACM Transactions on Audio, Speech, and Language Processing}, vol. 29, pp. 132--157, 2020.

\bibitem{straulino2023missing}
Elisa Straulino, Cristina Scarpazza, and Luisa Sartori,
\newblock ``What is missing in the study of emotion expression?,''
\newblock {\em Frontiers in Psychology}, vol. 14, pp. 1158136, 2023.

\bibitem{ladd2008intonational}
D~Robert Ladd,
\newblock {\em Intonational phonology},
\newblock Cambridge University Press, 2008.

\bibitem{russell1980circumplex}
James~A Russell,
\newblock ``A circumplex model of affect.,''
\newblock {\em Journal of personality and social psychology}, vol. 39, no. 6, pp. 1161, 1980.

\bibitem{nguyen2023expresso}
Tu~Anh Nguyen, Wei-Ning Hsu, Antony d'Avirro, Bowen Shi, Itai Gat, Maryam Fazel-Zarani, Tal Remez, Jade Copet, Gabriel Synnaeve, Michael Hassid, et~al.,
\newblock ``Expresso: A benchmark and analysis of discrete expressive speech resynthesis,''
\newblock {\em arXiv preprint arXiv:2308.05725}, 2023.

\bibitem{liu2024generative}
Rui Liu, Yifan Hu, Yi~Ren, Xiang Yin, and Haizhou Li,
\newblock ``Generative expressive conversational speech synthesis,''
\newblock in {\em Proceedings of the 32nd ACM International Conference on Multimedia}, 2024, pp. 4187--4196.

\bibitem{schuller2013computational}
Bj{\"o}rn Schuller and Anton Batliner,
\newblock {\em Computational paralinguistics: emotion, affect and personality in speech and language processing},
\newblock John Wiley \& Sons, 2013.

\bibitem{scherer2003vocal}
Klaus~R Scherer,
\newblock ``Vocal communication of emotion: A review of research paradigms,''
\newblock {\em Speech communication}, vol. 40, no. 1-2, pp. 227--256, 2003.

\bibitem{banse1996acoustic}
Rainer Banse and Klaus~R Scherer,
\newblock ``Acoustic profiles in vocal emotion expression.,''
\newblock {\em Journal of personality and social psychology}, vol. 70, no. 3, pp. 614, 1996.

\bibitem{gussenhoven2004phonology}
Carlos Gussenhoven,
\newblock ``The phonology of tone and intonation,''
\newblock 2004.

\bibitem{prieto2015intonational}
Pilar Prieto,
\newblock ``Intonational meaning,''
\newblock {\em Wiley Interdisciplinary Reviews: Cognitive Science}, vol. 6, no. 4, pp. 371--381, 2015.

\bibitem{shriberg2005spontaneous}
Elizabeth Shriberg,
\newblock ``Spontaneous speech: how people really talk and why engineers should care.,''
\newblock in {\em INTERSPEECH}, 2005, pp. 1781--1784.

\bibitem{de2015information}
Laura~E De~Ruiter,
\newblock ``Information status marking in spontaneous vs. read speech in story-telling tasks--evidence from intonation analysis using gtobi,''
\newblock {\em Journal of Phonetics}, vol. 48, pp. 29--44, 2015.

\bibitem{wu2025cnsced}
Xiaolong Wu, Chaobo Song, Shanshan Xiang, Ronghe Cao, Chang Feng, Hankiz Yilahun, Mingxing Xu, Askar Hamdulla, and Thomas~Fang Zheng,
\newblock ``A chinese natural speech complex emotion dataset based on emotion vector annotation method: X. wu et al.,''
\newblock {\em Language Resources and Evaluation}, pp. 1--22, 2025.

\bibitem{busso2008iemocap}
Carlos Busso, Murtaza Bulut, Chi-Chun Lee, Abe Kazemzadeh, Emily Mower, Samuel Kim, Jeannette~N Chang, Sungbok Lee, and Shrikanth~S Narayanan,
\newblock ``Iemocap: Interactive emotional dyadic motion capture database,''
\newblock {\em Language resources and evaluation}, vol. 42, no. 4, pp. 335--359, 2008.

\bibitem{Lameris2023Prosody}
Harm Lameris, Shivam Mehta, Gustav~Eje Henter, Joakim Gustafson, and {\'E}va Sz{\'e}kely,
\newblock ``Prosody-controllable spontaneous tts with neural hmms,''
\newblock in {\em ICASSP 2023-2023 IEEE International Conference on Acoustics, Speech and Signal Processing (ICASSP)}. IEEE, 2023, pp. 1--5.

\bibitem{matsunaga2022improving}
Yuta Matsunaga, Takaaki Saeki, Shinnosuke Takamichi, and Hiroshi Saruwatari,
\newblock ``Improving robustness of spontaneous speech synthesis with linguistic speech regularization and pseudo-filled-pause insertion,''
\newblock {\em arXiv preprint arXiv:2210.09815}, 2022.

\bibitem{Li2023Towards}
Weiqin Li, Shun Lei, Qiaochu Huang, Yixuan Zhou, Zhiyong Wu, Shiyin Kang, and Helen Meng,
\newblock ``Towards spontaneous style modeling with semi-supervised pre-training for conversational text-to-speech synthesis,''
\newblock {\em arXiv preprint arXiv:2308.16593}, 2023.

\bibitem{chen2016xgboost}
Tianqi Chen and Carlos Guestrin,
\newblock ``Xgboost: A scalable tree boosting system,''
\newblock in {\em Proceedings of the 22nd acm sigkdd international conference on knowledge discovery and data mining}, 2016, pp. 785--794.

\bibitem{conneau2020unsupervised}
Alexis Conneau, Alexei Baevski, Ronan Collobert, Abdelrahman Mohamed, and Michael Auli,
\newblock ``Unsupervised cross-lingual representation learning for speech recognition,''
\newblock {\em arXiv preprint arXiv:2006.13979}, 2020.

\bibitem{zhao2022m3ed}
Jinming Zhao, Tenggan Zhang, Jingwen Hu, Yuchen Liu, Qin Jin, Xinchao Wang, and Haizhou Li,
\newblock ``M3ed: Multi-modal multi-scene multi-label emotional dialogue database,''
\newblock {\em arXiv preprint arXiv:2205.10237}, 2022.

\bibitem{park2024let}
Se~Jin Park, Chae~Won Kim, Hyeongseop Rha, Minsu Kim, Joanna Hong, Jeong~Hun Yeo, and Yong~Man Ro,
\newblock ``Let's go real talk: Spoken dialogue model for face-to-face conversation,''
\newblock {\em arXiv preprint arXiv:2406.07867}, 2024.

\bibitem{zhao2025clearervoice}
Shengkui Zhao, Zexu Pan, and Bin Ma,
\newblock ``Clearervoice-studio: Bridging advanced speech processing research and practical deployment,''
\newblock {\em arXiv preprint arXiv:2506.19398}, 2025.

\bibitem{chen2025minmo}
Qian Chen, Yafeng Chen, Yanni Chen, Mengzhe Chen, Yingda Chen, Chong Deng, Zhihao Du, Ruize Gao, Changfeng Gao, Zhifu Gao, et~al.,
\newblock ``Minmo: A multimodal large language model for seamless voice interaction,''
\newblock {\em arXiv preprint arXiv:2501.06282}, 2025.

\bibitem{xu2025qwen2}
Jin Xu, Zhifang Guo, Jinzheng He, Hangrui Hu, Ting He, Shuai Bai, Keqin Chen, Jialin Wang, Yang Fan, Kai Dang, et~al.,
\newblock ``Qwen2. 5-omni technical report,''
\newblock {\em arXiv preprint arXiv:2503.20215}, 2025.

\end{thebibliography}

\end{document}